\documentclass{article}

\newcommand{\p}[1]{(\ref{#1})}

\newcommand{\hP}{{\hat P}}
\newcommand{\hQ}{{\hat Q}}
\newcommand{\hbQ}{{\hat{\bar Q}}}

\newcommand{\bD}{{\overline D}{}}
\newcommand{\bQ}{{\overline Q}{}}
\newcommand{\bS}{{\overline S}{}}
\newcommand{\bPsi}{{\overline \Psi}{}}
\newcommand{\bT}{{\overline T}}
\newcommand{\bV}{{\overline V}}
\newcommand{\bxi}{{\bar\xi}}
\newcommand{\bt}{{\bar\theta}}
\newcommand{\bpsi}{{\bar\psi}}

\newcommand{\be}{\begin{equation}}
\newcommand{\ee}{\end{equation}}
\newcommand{\bea}{\begin{eqnarray}}
\newcommand{\eea}{\end{eqnarray}}

\newcommand{\ba}{\begin{array}} \newcommand{\ea}{\end{array}}

\newcommand{\nn}{\nonumber}

\usepackage{amscd,amsmath,amssymb}

\def\theequation{\arabic{section}.\arabic{equation}}

\begin{document}
\thispagestyle{empty}
\vspace{2cm}
\begin{flushright}
%Draft 1 \\
%23.05.2009\\
\end{flushright}\vspace{2cm}
\begin{center}
{\Large\bf Potentials in N=4 superconformal mechanics}
\end{center}
\vspace{1cm}

\begin{center}
{\large\bf Stefano Bellucci${}^{a}$ and  Sergey Krivonos${}^{b}$ }
\end{center}

\begin{center}
${}^a$ {\it
INFN-Laboratori Nazionali di Frascati,
Via E. Fermi 40, 00044 Frascati, Italy} \vspace{0.2cm}

${}^b$ {\it
Bogoliubov  Laboratory of Theoretical Physics, JINR,
141980 Dubna, Russia} \vspace{0.2cm}

\end{center}
\vspace{2cm}

\begin{abstract}\noindent
Proceeding from nonlinear realizations of (super)conformal
symmetries, we explicitly demonstrate that adding the harmonic
oscillator potential to the action of conformal mechanics
does not break these symmetries but modifies
the transformation properties of the (super)fields. We also
analyze the possibility to introduce potentials in $N=4$
supersymmetric mechanics by coupling it with auxiliary fermionic
superfields. The new coupling we considered does not introduce new
fermionic degrees of freedom - all our additional fermions are
purely auxiliary ones. The new bosonic components have a
first order kinetic term and therefore they serve as spin degrees
of freedom. The resulting system contains, besides the potential
term in the bosonic sector, a non-trivial spin-like interaction in
the fermionic sector. The superconformal mechanics we constructed
in this paper is invariant under the full $D(2,1;\alpha)$
superconformal group. This invariance is not evident and is
achieved within  modified  (super)conformal transformations of the
superfields.
\end{abstract}

\newpage
\setcounter{page}{1}
\section{Introduction}
The conformal symmetry in one dimension is so powerful, that in
the one particle case it completely fixes the potential to be
$1/x^2$ \cite{AFF}. The supersymmetric extensions of conformal
mechanics \cite{n2,IKL1} add to the theory some fermions
interacting with the single bosonic field $x$, but in the bosonic
sector the potential is still  the same. So it seems that the
question whether we can add something to the bosonic potential
without breaking (super)conformal invariance has a unique answer -
no. Nevertheless, it is not completely true. Indeed, the standard
description of the one dimensional conformal invariance consists
in the statement that the Hamiltonian \be\label{i1} H= \frac{1}{2}
p^2 +\frac{g^2}{2 x^2} \ee forms the $so(1,2)$ algebra together
with the generators of the dilatation $D$ and conformal boost $K$
defined as \be\label{i2} D=xp, \qquad K= x^2, \ee with respect to
the canonical Poisson bracket \be\label{i3} \left\{ x,p\right\}=1.
\ee Now it is completely clear that the modified Hamiltonian $\hat
H$ \be\label{i4} \hat H = H+m^2K \ee will do the same job
perfectly forming the same $so(1,2)$ algebra with generators $D$
and $K$ \p{i2}. Thus, the additional harmonic oscillator potential
is also admissible without breaking conformal symmetry.   It has
been firstly shown in \cite{AP1} that this additional harmonic
oscillator potential just modifies  the realization of the
conformal group, keeping the resulting action invariant under
conformal group transformations. Unfortunately, the approach
presented in \cite{AP1} is not fully appropriate and the nice idea
of the modification of conformal group transformations could not
be immediately extended to the $N=2,4$ supersymmetric cases. In
the present paper (Sections 2-4) we will demonstrate that this
additional harmonic oscillator term can be easily obtained within
the nonlinear realizations framework in which the conformal mechanics (together with
its $N=2,4$ superextensions) describes the motion along geodesics
in the group space of the $d=1$ (super)conformal group
\cite{IKL0,IKL1}. In such approach, the almost trivial bosonic case
(Section 2) has natural and straightforward $N=2$ (Section 3) and
$N=4$ (Section 4) supersymmetric analogs.

Another issue we analyzed in this paper is the generation of
bosonic potentials in $N=4$ supersymmetric mechanics through
coupling it with the auxiliary fermionic supermultiplet. This
additional supermultiplet enters the action in a rather special
manner:
\begin{itemize}
\item the fermionic components $\psi^a, \bar\psi_a$ appear in the action only hatted by the time derivative.
Thus the corresponding equations of motion are nothing but the
conservation laws, and all fermionic components could be expressed
in terms of the remaining components;
\item  the bosonic components $w^i, {\bar w}_i$ have a first order kinetic term and therefore
serve as spin degrees of freedom;
\item  the coupling
constant $g$ in the bosonic potential is the square of the norm of
the spin degrees of freedom $w^i: g=w^i{\bar w}_i=const$.
\end{itemize}
The idea to generate the bosonic potential by coupling with
additional supermultiplets has been firstly proposed in \cite{DI}.
In this paper the Authors introduced coupling with a
supermultiplet containing physical fermions, thus finishing with a
system with a doubled number of fermions, in contrast with our
case where all fermions are auxiliary. Our approach in this paper
is very close to those one recently proposed in \cite{IFL}. The
component actions we constructed in this paper have to
coincide with the ones from \cite{IFL}, because the main ingredient
- the action describing the coupling of the basic $(1,4,3)$
supermultiplet with the auxiliary fermionic $(0,4,4)$ one - is
unique and is completely fixed by $N=4$ Poincar\'{e} supersymmetry
(up to an overall constant).

\section{Bosonic case: justification of the idea}
The standard conformal algebra in $d=1$ is the $so(1,2)$ one
spanned by the generators of translations ($P$), dilatations ($D$)
and conformal boosts ($K$) \p{alg1}. One of the simplest ways to
construct a one-dimensional system which is conformally invariant
is to use the method of covariant reduction \cite{IKL0}. In
application to this simplest case this method includes the
following steps:
\begin{itemize}
\item realization of the conformal group $SO(1,2)$ in some coset;
\item building the Cartan forms;
\item imposing invariant constraints on the Cartan forms which result in the desired equations
of motion.
\end{itemize}
Let us choose the following parametrization of the $SO(1,2)$ group
space
\be\label{g1}
g = e^{itP}\; e^{izK}\; e^{iuD}
\ee
where the
coordinates $u$ and $z$ are functions of the time variable $t$.
Thus, we are dealing with the nonlinear realization of $SO(1,2)$ in
its group space. The Cartan forms for the group element $g$ \p{g1}
read \be\label{CF0} g^{-1}\; dg = i\omega_P P + i\omega_K K +
i\omega_D D \;, \ee where \be\label{CF1} \omega_P=e^{-u} dt, \quad
\omega_D = du-2z dt,\quad \omega_K=e^u\left[ dz +z^2 dt\right].
\ee In the present case the coset coincides with the group space,
therefore all Cartan forms \p{CF1} are invariant under $SO(1,2)$
transformations realized as left multiplications of the group
element $g$ \p{g1}. The set of constraints we are going to impose
on the forms \p{CF1} reads \cite{IKL1} \be\label{con1} \omega_D=0,
\qquad \omega_K=g^2 \omega_P , \ee where $g$ is a free parameter
with the dimension of the mass. The first constraint in \p{con1}
\be\label{IH1} \omega_D = du-2z dt=0 \quad \Rightarrow \quad
z=\frac{1}{2}\dot{u} \ee is just a simplest version of the Inverse
Higgs phenomenon \cite{IO}. Its meaning is rather simple -- we do
not need an independent field $z(t)$ in order to realize the
conformal invariance in the group space. Instead we may use the
time derivative of the dilaton $\dot{u}$ which has the same
transformation properties with respect to the $SO(1,2)$ group.

The second constraint in \p{con1} is a dynamical one. Using the
\p{IH1} it may be rewritten in the familiar form as
\be\label{conf} \ddot{x}=\frac{g^2}{x^3}, \qquad \mbox{where}
\qquad x\equiv e^{\frac{u}{2}}. \ee Clearly, the equation of
motion \p{conf} follows from the action of conformal mechanics
\cite{AFF} \be\label{L1} S=\int dt \left( \frac{ {\dot{u}}^2}{2}-
\frac{g^2}{2 x^2}\right). \ee It is not unexpected now that the
action \p{L1} is invariant under conformal transformations
\be\label{transf1} \delta t = f(t) = a+bt+ct^2, \delta u =
\dot{f}. \ee Let us stress that the explicit form of \p{transf1}
simply follows from the left action of the conformal group
$SO(1,2)$ on the group element $g$ \p{g1}.

All this is not new and has been known for a long time. In order
to learn something new, let us change the parametrization \p{g1}
and associate the time variable $t$ with the generator $P+m^2K$,
where $m$ is an additional parameter with the dimension of the
mass as \be\label{g2} {\tilde g} = e^{it\left( P +m^2 K\right)}\;
e^{izK}\; e^{iuD} \;. \ee The explicit relation between the coset
element $\tilde g$ \p{g2} and $g$ \p{g1} is given by
\be\label{g1g2} e^{it\left(P+m^2 K\right)}e^{izK} e^{iuD}=e^{i
\frac{\tan[m t]}{m}P}e^{i\left(z\cos^2[m t]+ m \cos[m t]\sin[m
t]\right)K} e^{i\left( u-2\log[\cos[m t]]\right)D}. \ee The Cartan
forms \p{CF0} are slightly changed to be \be\label{CF}
\widetilde{\omega}_P=e^{-u} dt, \quad \widetilde{\omega}_D = du-2z
dt,\quad \widetilde{\omega}_K=e^u\left[ dz +z^2 dt +m^2 dt\right].
\ee Now we impose the same constraints as before \p{con1}. As a
result, we are ending up with the following equation of motion:
\be\label{conf1} \ddot{x}=-m^2 x+\frac{g^2}{x^3}, \qquad
\mbox{where again} \qquad x= e^{\frac{u}{2}}. \ee Clearly, this
equation follows from the following action: \be\label{L2}
\tilde{S}=\int dt \left[ \frac{ {\dot{x}}^2}{2}- \frac{m^2 x^2}{2}
-\frac{g^2}{2 x^2}\right]. \ee Thus, by construction, the action
\p{L2}, which describes the conformal mechanics equipped with an
additional harmonic oscillator term, has to be invariant under the
"conformal" group $SO(1,2)$! It is not too hard to find the
corresponding realization of this symmetry
\be\label{transf2}
\delta t = \tilde{f}(t)=a\left( 1+ Cos(2m t)\right)+
\frac{b}{2m}Sin(2m t)+\frac{c}{2m^2} \left(1- Cos(2m t)\right)
,\quad \delta u=\dot{\tilde{f}},
\ee
where the parameters
$(a,b,c)$ are, as before, the parameters for translations,
conformal boosts and dilatations, respectively. The function
${\tilde f}(t)$ which collects all parameters of the $SO(1,2)$
transformations obeys, in view of \p{transf2}, the constraint
\be\label{f1}
\frac{d}{dt}\left[ \ddot{\tilde f}+ 4 m^2 {\tilde f} \right] =0.
\ee
Thus, the conformal mechanics with the added harmonic oscillator
term (1-particle Calogero-Moser system) is invariant under the
$SO(1,2)$ group which has a non-canonical realization on the time
variable $t$.

Two comments have to be added here. Firstly, in the limit
$m\rightarrow 0$ the transformations \p{transf2} are reduced to
the standard ones \p{transf1}, as it should be. Secondly, the
kinetic term in the action \p{L2} is invariant under \p{transf2}
only together with the oscillator term, while the conformal
potential is invariant by itself.

The invariance of the action \p{L2} with respect the $SO(1,2)$
transformations \p{transf2} has been firstly demonstrated in
\cite{AP1}. Although we do not use the realization of the Virasoro
group in the corresponding coset, which was the key ingredient in
\cite{AP1}, it makes sense to consider the present approach as a
further development of the nice ideas of this paper.

In the next Sections we will extend our description to the $N=2$ and $N=4$ supersymmetric cases.

\setcounter{equation}0
\section{N=2 superconformal mechanics}
The simplest nontrivial extension of the conformal mechanics
corresponds to the $N=2$ case with $SU(1,1|1)$ superconformal
group \cite{n2}. The geometric construction of $N=2$
superconformal mechanics within the framework of the nonlinear
realization of $SU(1,1|1)$ in the coset $SU(1,1|1)/U(1)$ has been
carried out in \cite{IKL1}. Without deeply going in the details,
our extension of the consideration in \cite{IKL1} looks as
follows. As usual, we are starting from the $N=2$ superconformal
group $SU(1,1|1)$. Its superalgebra includes  the four bosonic
generators $\left\{ P,D,K, V_3\right\}$ and the four fermionic
ones $\left\{ Q^1,\bQ_1,S^1, \bS_1\right\}$. Their commutators
follow from the general formulas given in Appendix
\p{alg1}-\p{alg4} with $\alpha=-1$.

We will realize this group in the coset $SU(1,1|1)/U(1)$
parameterized as \be\label{g3} g=e^{it\hP}e^{\theta \hQ +\bt
\hbQ}e^{izK}e^{\psi S^1+\bpsi \bS_1}e^{iuD}, \ee where
\be\label{def1} \hP=P+2im V_3 +m^2 K,\quad \hQ=Q^1+i m S^1, \quad
\hbQ=\bQ_1-i m \bS_1, \ee and all the coordinates $(u,z,
\psi,\bpsi)$ are $N=2$ superfields which depend on $(t,
\theta,\bt)$. One should stress that the $U(1)$ generator is
anti-hermitian \p{conjug}, so the operator $\hP$ \p{def1} is
hermitian. The only difference with the presentation in
\cite{IKL1} is given by the $m$-dependent terms in
\p{g3},\p{def1}.

In what follows we shall need the explicit structures of several Cartan forms in the expansion
$g^{-1}dg$ over the generators,
\bea\label{CF2}
&& \omega_P=e^{-u}\left( dt +i\theta d\bt+i \bt d\theta\right) \equiv e^{-u} d\tilde{t},\quad
\omega_{Q^1}=e^{-\frac{1}{2}u}\left( d\theta -\psi d\tilde{t}\right), \nonumber \\
&& \omega_D= du-2xd\tilde{t}-2id\bt\psi -2id\theta\bpsi, \nonumber \\
&&\omega_{S^1}=e^{\frac{1}{2}u}\left( d\psi -zd\theta -i \psi\bpsi
d\theta +(z+i m)\psi d\tilde{t}+im d\theta\right). \eea The
constraints we impose on the Cartan forms are the same as in
\cite{IKL1} \be\label{con3} \omega_D=0, \qquad \omega_{S^1} =i g
\omega_{Q^1}, \ee where, as before, the arbitrary parameter $g$
has the dimension of the mass. As a result we will get two sets of
equations which follow from \p{con3} \be\label{ih2}
z=\frac{1}{2}\dot{u}, \quad \psi=-\frac{i}{2}\bD u,\;
\bpsi=-\frac{i}{2} Du, \ee and \be\label{n2cal} \left[ D,
\bD\right] X = 2m X +\frac{2g}{X}, \qquad X \equiv
e^{\frac{1}{2}u}, \ee where the semi-covariant (fully invariant
only under Poincar\'{e} supersymmetry) spinor derivatives are
defined by \be D=\frac{\partial}{\partial\theta}+i\bt\partial_t,\;
\bD=\frac{\partial}{\partial\bt}+i\theta\partial_t, \qquad
\left\{D,\bD \right\}=2i\partial_t . \ee The equations \p{ih2}
express unessential Goldstone superfields $(z,\psi,\bpsi)$ in
terms of the super-dilaton $u$, while the equation \p{n2cal} is
the dynamical one. Clearly, it can be obtained from the following
superfield action: \be\label{L3} S=\int dt d^2\theta \left[ D X
\bD X +m X^2 + 2 g \log(X) \right]. \ee It is worth to note that
the action \p{L3}, despite the presence of the harmonic oscillator
potential $m X^2$, has to be invariant under the full $SU(1,1|1)$
superconformal group. In order to clarify this point, let us write
the transformations as follows: \be\label{realiz1} \delta
t=E-\frac{1}{2}\bt\; \bD E -\frac{1}{2} \theta D E, \quad
\delta\theta=-\frac{i}{2}\bD E,\; \delta \bt =-\frac{i}{2} D
E,\qquad \delta u = \dot{E}, \ee where $E$ is a superfunction
collecting all parameters of the $SU(1,1|1)$ group. In the
standard realization (with $m=0$ ) this function obeys the
following constraint: \cite{IKL1} \be \frac{\partial}{\partial t}
\left[ D,\bD\right] E =0, \ee which leaves in $E$ just the
parameters of the $SU(1,1|1)$ group. In the present case this
constraint is modified to become \be\label{eqq}
\frac{\partial}{\partial t} \left[ D,\bD\right] E =4m \dot{E}. \ee
One may easily check that the solution of the modified equation
\p{eqq} also contains the set of parameters corresponding to the
transformations of the $SU(1,1|1)$ group. Nevertheless,  the
realization of this group on the superspace $(t,\theta,\bt)$ and
superfield $u$ \p{realiz1} is different now. The action \p{L3} is
invariant with respect to these transformations. In a full analogy
with the bosonic case, the term $g \log(X)$ is invariant by
itself, while the kinetic term is invariant only together with the
harmonic potential. Thus, we explicitly demonstrated that the
$N=2$ superconformal mechanics, being equipped with the harmonic
oscillator potential, admits the same invariance with respect to
$N=2$ superconformal symmetry $SU(1,1|1)$, but with a different
realization.

To conclude, it makes sense to note that in principle one could
ignore the coset construction and just ask which invariance does
the superfield action \p{L3} possess. Looking for the answer one
may write the general transformations in the form \p{realiz1} and
then immediately get the constraint \p{eqq}  on the superfunction
$E$ which selects just the $SU(1,1|1)$ group.
\setcounter{equation}0
\section{N=4 superconformal mechanics}
The extension of our previous consideration to the case of $N=4$
superconformal mechanics goes almost straightforwardly. We will
start with the $su(1,1|2)$ superalgebra which is isomorphic to
$D(2,1;-1)$ (see Appendix). Next, similarly to \cite{IKL1}, we
will realize $SU(1,1|2)$ in the coset  $SU(1,1|2)/SU(2)$
parameterized as \be\label{g4} g=e^{it\hP}e^{\theta_i \hQ{}^i
+\bt{}^i \hbQ{}_i}e^{izK}e^{\psi_i S^i+\bpsi{}^i \bS_i}e^{iuD},
\ee where \be\label{def41} \hP=P+2im V_3 +m^2 K,\quad
\hQ{}^1=Q^1+i m S^1,\hQ{}^2=Q^2-i m S^2, \quad \hbQ_1=\bQ_1-i m
\bS_1,\hbQ_2=\bQ_2+i m \bS_2 \ee and all the coordinates $(u,z,
\psi,\bpsi)$ are now $N=4$ superfields which depend on $(t,
\theta_i,\bt{}^i)$. With our choice \p{def41} the generators
$\left\{\hP,\hQ{}^i,\hbQ_i\right\}$ span the $N=4, d=1$ super
Poincar\'{e} algebra.

In order to construct the equations of motion, we have to impose
covariant constraints on the Cartan forms. We will choose the same
constraints as in \cite{IKL1}, namely \be\label{conN4} \omega_D=0,
\qquad \omega_{S^i} =i g \omega_{Q^i}. \ee Using explicit
expressions for the Cartan forms \bea
\omega_D &=&  du -2z d{\tilde t}-2i d\bt{}^i\xi_i -2id\theta_i \bxi{}^i,\qquad \omega_{Q^i}=e^{-\frac{u}{2}}d\theta_i + d{\tilde t}\left(\ldots\right),\nn \\
\omega_{S^1} &=& e^{\frac{u}{2}} \left[ d\xi_1-i\xi_i \bxi{}^i d\theta_1 -2i\xi_1 \xi_2d\bt{}^2+ \left(i m-z\right)d\theta_1\right]+
d{\tilde t}\left(\ldots\right),\nn\\
\omega_{S^2} &=& e^{\frac{u}{2}}\left[ d\xi_2-i\xi_i \bxi{}^i
d\theta_2 +2i\xi_1 \xi_2d\bt{}^1- \left(i
m+z\right)d\theta_2\right]+ d{\tilde t}\left(\ldots\right), \eea
where the covariant differential of $t$ is defined as \be d{\tilde
t} = dt -i \left( d \bt{}^i \theta_i+d\theta_i \bt{}^i\right), \ee
we will get the following set of equations from \p{conN4}: \bea
&& z=\frac{1}{2}{\dot u}, \qquad \xi_i=-\frac{i}{2} \bD_i u, \; \bxi{}^i =-\frac{i}{2} D^i u, \label{ih4} \\
&& D^i D_i e^u=0, \; \bD_i \bD{}^i e^u=0, \qquad \left[D^i ,\bD_i\right] e^u =-8g, \label{eq41} \\
&& D^1 \bD_2 u= D^2 \bD_1 u=0, \qquad \left( D^1 \bD_1 -D^2 \bD_2 \right) u = 4m \label{eq42}.
\eea
Here, we introduced the spinor covariant derivatives as
\be
D^i=\frac{\partial}{\partial \theta_i}+i\bt{}^i\partial_t,\;
\bD^i=\frac{\partial}{\partial \bt{}^i}+i\theta_i\partial_t,\qquad \left\{ D^i,\bD_j\right\}=2i\partial_t.
\ee
The meaning of the equations \p{ih4}-\p{eq42} is clear:
\begin{itemize}
\item the equations \p{ih4} express unessential superfields
$\left\{z,\xi_i,\bxi^i\right\}$ in terms of the superdilaton $u$;
\item the constraints \p{eq41} are off-shell irreducibility conditions: they reduce
the components content of the $N=4$ superfield $u$ to 1 physical
and 3 auxiliary bosonic fields and four fermionic fields
\cite{IKL1};
\item equations \p{eq42} are dynamical: they serve to eliminate the triplet of auxiliary fields
and give rise to equations of motion
for the physical fields.
\end{itemize}
The component action has a very simple form \be S=\int dt\left[
\frac{{\dot y}^2}{2} +\frac{i}{2}\left( \bpsi_i{\dot\psi}{}^i
-{\dot\bpsi}_i\psi^i\right)-\frac{m^2 y^2}{2}-\frac{g^2}{2y^2}
+m\bpsi_i \psi^i+ \frac{g}{y^2}\left(
\bpsi_1\psi^1-\bpsi_2\psi^2\right)+3\frac{\bpsi_1\psi^1\bpsi_2\psi_2}{y^2}\right]
\ee where \be\label{ac4} y=e^{\frac{u}{2}}|_{\theta=\bt=0},
\psi^1=D^1 e^{\frac{u}{2}}|_{\theta=\bt=0},\psi^2=\bD_2
e^{\frac{u}{2}}|_{\theta=\bt=0}. \ee By construction the action
\p{ac4} is invariant with respect to the $su(1,1|2)$ superalgebra
realized in the modified coset \p{g4}.

\setcounter{equation}0
\section{Potentials in N=4 superconformal mechanics}
One of the most restrictive features of $N=4$ supersymmetric
mechanics based on the $(1,4,3)$ supermultiplet is a specific
generation of potential terms from constants in the defining
constraints. It has been shown many years ago \cite{IKL1} that the
constraints defining the irreducible $N=4, d=1$ supermultiplet
with $(1,4,3)$ components content are uniquely  fixed to be
\be\label{la1} D^i D_i X =  g f,\; \bD_i \bD{}^i X = g {\bar f},
\qquad  \left[ D^i, \bD_i\right] X=2 g c, \ee where the set of
constants $f, {\bar f}, c$ are related by \be\label{la2}
c^2+f{\bar f}=1. \ee Clearly, one may always choose \be c=0,\;
f=-{\bar f}=i \ee to have \be\label{la3} D^i D_i X =  i g ,\;
\bD_i \bD{}^i X = -i g , \qquad  \left[ D^i, \bD_i\right] X=0. \ee
Now, the most general action of the one-particle $N=4$
supersymmetric mechanics reads \be\label{la4} S=-\int dt d^4\theta
{\cal F}(X), \ee with ${\cal F}(X)$ being an arbitrary function of
the superfield $X$. It is not hard to get the bosonic part of the
component action (with auxiliary fields excluded by their
equations of motion) \be\label{la5} S_B\sim \int dt \left[ F''
{\dot x}{}^2 + g^2 F''\right], \qquad x\equiv X|_{\theta=\bt=0},
F(x)\equiv {\cal F}(X)|_{\theta=\bt=0}. \ee Finally, one could
bring the kinetic term to the flat one \be\label{la6} S_B \sim
\int dt \left[ {\dot y}{}^2+g^2 (y')^2\right], \qquad F''=(y')^2,
\ee where  $y'(y)$ is considered as a function of $y$. The
additional requirement of conformal invariance, i.e. $y'\sim 1/y$,
completely fixes everything, unambiguously restoring ${\cal F}(X)
\sim X \log X$. Thus, everything is strictly fixed by $N=4$
superconformal invariance ($SU(1,1|2)$ in the case at hands).

Another possibility to get the potential term in $N=4$
supersymmetric mechanics has been proposed in \cite{DI}. The main
idea was to couple the $(1,4,3)$ supermultiplet to the fermionic
$(0,4,4)$ one. The price one has to pay for this is a doubled up
number of physical fermions in the resulting system. Here we will
use the same idea of coupling but with a different action. In our
action the fermions appear only through the time derivatives which
can be replaced, without breaking of supersymmetry, by  auxiliary
fermions. Moreover, the bosonic fields in the action have the
kinetic term which is linear in the time derivatives, and
therefore these bosonic fields acquire the interpretation of spin
degrees of freedom.

Our starting point is the $(1,4,3)$ supermultiplet $X$ with $g=0$
\p{la3} and the fermionic $(0,4,4)$ supermultiplet
$\Psi{}^a,\bPsi_a$ defined by the constraints\footnote{If we
combine the spinor derivatives $D^i,\bD^i$ in the quartet of
spinor derivatives $\nabla^{ia}=\left\{ D^i, \bD^i\right\}$ then
the constraints \p{la7} acquire the familiar form
$\nabla^{i(a}\Psi^{b)}=0$}. \be\label{la7} D^i \Psi{}^1=0, \; D^i
\Psi{}^2+\bD^i \Psi{}^1=0, \; \bD_i \Psi{}^2=0. \ee We introduce
the coupling of these supermultiplets by considering the following
action: \be\label{la8} S=S_1 +S_2 =-\frac{1}{32}\int dt d^4\theta
{\cal F}(X)-\frac{1}{32}\int dt d^4\theta X \Psi^a \bPsi_a . \ee
After integration over theta's, the components action which
follows from \p{la8} reads \bea\label{la9} S&=&\int
dt\left[\frac{1}{8}F'' {\dot x}{}^2-\frac{1}{16} F''
A^{ij}A_{ij}+\frac{i}{8} F''\left(\dot\eta{}^i\bar\eta_i-
\eta^i\dot{\bar\eta}_i\right)+\frac{1}{8}F'''\eta^i\bar\eta{}^jA_{ij}-\frac{1}{32}F^{(4)}\eta^i\eta_i\bar\eta_j\bar\eta{}^j\right]+ \nn\\
&&\int dt\left[ -x \left({\dot \psi}{}^1{\dot\bpsi}{}^2-{\dot \psi}{}^2{\dot\bpsi}{}^1\right)+\frac{i}{4} x \left( {\dot v}_i {\bar v}{}^i-
v_i\dot{\bar v}{}^i\right)+\frac{1}{4}A_{ij}v^i{\bar v}{}^j+\right.\nn\\
&& \left. \frac{1}{2}\eta_i\left({\bar v}{}^i \dot\bpsi{}^2+v^i \dot\psi{}^2\right)+\frac{1}{2}\bar\eta{}^i\left(v_i \dot\psi{}^1+{\bar v}_i\dot\bpsi{}^1\right)
\right],
\eea
where
\bea
&& x\equiv X|,\; A_{(ij)}\equiv \frac{1}{2}\left[ D_i,\bD_j\right] X|,\qquad \eta^i\equiv -iD^i X|,\;
\bar\eta{}_i\equiv -i\bD_i X|, \nn\\
&& \psi^a\equiv \Psi{}^a|,\qquad v^i\equiv -D{}^i\bPsi{}^2|, \;
{\bar v}_i\equiv  \bD_i \Psi{}^1|, \eea and, as usual, $|$ in the
r.h.s. denotes the $\theta=\bt=0$ limit. In the component action
\p{la9} the fermionic fields $\psi^a$ and $\bpsi{}^a$ enter only
through the time derivatives. Let us replace these time
derivatives by new fermionic fields $\xi^a$ and $\bxi{}^a$ as
\be\label{xi} \xi^a= {\dot\psi}{}^a, \qquad \bxi^a=\dot\bpsi{}^a.
\ee This is nothing but the reduction from the $(0,4,4)$
supermultiplet to the auxiliary $(4,4,0)$ one
\cite{GR,root,SM}\footnote{The superfield version of such a reduction
reads $V^i=\nabla^{ia} \Psi_a$, where the $N=4$ superfields $V^i$
start with $v^i$ and, by construction, obey the constraints
$\nabla^{a(i}V^{j)}=0$.}. This reduction is compatible with $N=4$
supersymmetry. Indeed, the components of the $\Psi^a$ have the
following transformation properties under $N=4$ Poincar\'{e}
supersymmetry \be\label{ad1} \delta \psi^1=-\bar\epsilon{}^i {\bar
v}_i,\; \delta\psi^2=\epsilon_i{\bar v}{}^i,\quad \delta
v^i=-2i\epsilon^i\dot\bpsi{}^1+2i\bar\epsilon{}^i\dot\bpsi{}^2,\;
\delta {\bar
v}_i=-2i\epsilon_i\dot\psi{}^1+2i\bar\epsilon_i\dot\psi{}^2. \ee
{}From \p{ad1} we learn the transformation properties of the new
fermions $\xi^a,\bxi{}^a$ \be\label{ad2} \delta
\xi^1=-\bar\epsilon{}^i \dot{\bar v}_i,\;
\delta\xi^2=\epsilon_i\dot{\bar v}{}^i,\quad \delta
v^i=-2i\epsilon^i\bxi{}^1+2i\bar\epsilon{}^i\bxi{}^2,\; \delta
{\bar v}_i=-2i\epsilon_i\xi{}^1+2i\bar\epsilon_i\xi{}^2. \ee Now
one may easily check that the action \bea\label{ad3} S&=&\int
dt\left[\frac{1}{8}F'' {\dot x}{}^2-\frac{1}{16} F''
A^{ij}A_{ij}+\frac{i}{8} F''\left(\dot\eta{}^i\bar\eta_i-
\eta^i\dot{\bar\eta}_i\right)+\frac{1}{8}F'''\eta^i\bar\eta{}^jA_{ij}-\frac{1}{32}F^{(4)}\eta^i\eta_i\bar\eta_j\bar\eta{}^j\right]+ \nn\\
&&\int dt\left[ -x \left({ \xi}{}^1{\bxi}{}^2-{ \xi}{}^2{\bxi}{}^1\right)+\frac{i}{4} x \left( {\dot v}_i {\bar v}{}^i-
v_i\dot{\bar v}{}^i\right)+\frac{1}{4}A_{ij}v^i{\bar v}{}^j+\right.\nn\\
&& \left. \frac{1}{2}\eta_i\left({\bar v}{}^i \bxi{}^2+v^i
\xi{}^2\right)+\frac{1}{2}\bar\eta{}^i\left(v_i \xi{}^1+{\bar
v}_i\bxi{}^1\right) \right], \eea is invariant under \p{ad2},
provided the components of the $X$ supermultiplet transform in a
standard way as\footnote{We defined symmetrization over indices as
$a_{(ij)}\equiv \frac{1}{2}\left( a_{ij}+a_{ji}\right)$.}
\be\label{ad4} \delta
x=-i\epsilon_i\eta^i-i\bar\epsilon{}^i\bar\eta_i,\quad
\delta\eta{}^i=-\bar\epsilon{}^i{\dot x}-i\bar\epsilon{}^j
A^i_j,\; \delta \bar\eta_i=-\epsilon_i{\dot x}+i\epsilon_j
A_i^j,\quad \delta A_{ij} = -\epsilon_{(i}\dot\eta_{j)} +
\bar\epsilon_{(i}\dot{\bar\eta}{}_{j)}. \ee In the action \p{ad3}
the fields $\xi^a,\bxi{}^a$ and $A_{ij}$ are auxiliary ones.
Eliminating them by their equations of motion \be\label{ad5}
\xi^1=\frac{1}{2x}\eta_i{\bar v}{}^i, \quad
\xi^2=-\frac{1}{2x}\bar\eta{}^i{\bar v}_i, \qquad
A_{ij}=\frac{F'''}{F''}\eta_{(i}\bar\eta_{j)}+\frac{2}{F''}v_{(i}{\bar
v}_{j)}, \ee we will get the following action: \bea\label{la10}
S&=&\int dt\left[\frac{1}{8}F'' {\dot x}{}^2+\frac{i}{8}
F''\left(\dot\eta{}^i\bar\eta_i-
\eta^i\dot{\bar\eta}_i\right)+\frac{1}{32}\left(
\frac{3}{2}\frac{(F''')^2}{F''}-F^{(4)}\right)\eta^i\eta_i\bar\eta_j\bar\eta{}^j+
\frac{i}{4} \left( {\dot w}_i {\bar w}{}^i- w_i\dot{\bar
w}{}^i\right)+\right.
\nn\\
&& \left. \frac{1}{8x}\left(\frac{F'''}{ F''}+\frac{2}{x}\right)
\eta_i\bar\eta_j\left( w^i {\bar w}{}^j+ w^j{\bar
w}{}^i\right)-\frac{1}{8F'' x^2}\left(w^i{\bar w}_i\right)^2
\right], \eea where, in order to bring the kinetic term for $v^i$
to the standard form we introduce the new fields $w^i$ as \be
w^i=\frac{1}{\sqrt{x}}v^i. \ee Thus, we see that from our
fermionic superfields $\Psi{}^a$, there survived only bosonic
components $w^i,{\bar w}_i$ which enter the Lagrangian only
through first time-derivatives. After quantization these variables
become purely internal spin degrees of freedom. Moreover, from the
equations of motion for $w^i,{\bar w}_i$ one may conclude that \be
w^i{\bar w}_i=g=\mbox{ constant}, \ee and therefore, the last term
in \p{la10} is just the bosonic potential for $x$ \be\label{la11}
V_B=\frac{g^2}{8F'' x^2}. \ee Thus, we conclude that indeed by
coupling our $(1,4,3)$ superfield $X$ with the fermionic $(0,4,4)$
auxiliary supermultiplet $\Psi^a$ one may generate the bosonic
potential for the physical bosonic field $x$, together with terms
describing the spin interaction of the fermionic components
$\eta^i,\bar\eta_i$.

It is quite interesting to understand whether the action \p{la8}
could possess any type of $N=4$  superconformal symmetry. The key
point is to achieve the invariance of the second term $S_2$ in the
action \p{la8}, because the first one $S_1$ can be always chosen
to be superconformally invariant. Indeed, the superfield $X$ obeys
the constraints \p{la3} with $g=0$. The invariance of these
constraints under the $D(2,1;\alpha)$ group forces $X$  to
transform as \cite{IKLe1} \be\label{la13} \delta X = 2i\alpha
\left(\epsilon_i \bt{}^i+\bar\epsilon{}^i\theta_i\right)X \ee
while the superspace measure transformation reads \be\label{la14}
\delta dt d^4\theta = 2i \left(\epsilon_i
\bt{}^i+\bar\epsilon{}^i\theta_i\right) dt d^4\theta. \ee Clearly,
the superconformally invariant action for the supermultiplet $X$
has the following form: \be\label{ad5}
S_1^{Conf}=-\frac{1}{32}\int dt d^4\theta
\left(X\right)^{-\frac{1}{\alpha}}, \qquad \alpha \neq -1 \ee or
\cite{IKL1} \be\label{ad6} S_1^{Conf}=-\frac{1}{32}\int dt
d^4\theta X \log X, \qquad \alpha = -1. \ee The invariance of the
second term $S_2$ needs to be considered more carefully. First of
all, one may check that the action $S_2$ in the form of \p{ad3} is
invariant under the $D(2,1;\alpha)$ group for arbitrary $\alpha$,
provided the components transform under conformal supersymmetry as
\bea && \delta
v^i=-2it\left(\varepsilon^i\bxi{}^1-\bar\varepsilon{}^i\bxi{}^2\right),\qquad
\delta\bxi{}^1=-\alpha\bar\varepsilon_i v^i
+t\bar\varepsilon_i{\dot v}{}^i,\;
\delta\bxi{}^2=-\alpha\varepsilon_i v^i +t\varepsilon_i{\dot v}{}^i,\label{ad7a}\\
&& \delta x =-t\left(\varepsilon_i \eta^i+\bar\varepsilon{}^i\bar\eta{}_i\right),\qquad
\delta \eta^i=-2\alpha \bar\varepsilon{}^ix -\bar\varepsilon{}^i t {\dot x}-i t \bar\varepsilon{}^j A^i_j, \nn \\
&& \delta A_{ij}=-2(1+2\alpha )\left(
\varepsilon_{(i}\eta_{j)}-\bar\varepsilon{}_{(i}\bar\eta{}_{j)}\right)-
2t\left(
\varepsilon{}_{(i}\dot\eta{}_{j)}-\bar\varepsilon{}_{(i}\dot{\bar\eta}{}_{j)}\right).\label{ad7b}
\eea Thus, the action \p{ad3} with the properly chosen
superpotential $F$ as in \p{ad5}, \p{ad5} is invariant with
respect to the full $N=4$ superconformal group  $D(2,1;\alpha)$.

The crucial point in proving the superconformal invariance of our
action, was its form \p{ad3} obtained {\it after} reduction
\p{xi}. If we instead would check the invariance of the action
\p{la8} and limit ourselves by considering the {\it local}
transformations of fermionic superfields $\Psi^a$  we would get
that \be\label{la12} \delta\left(
\Psi^a\bPsi_a\right)=2i(1+\alpha) \left(\epsilon_i
\bt{}^i+\bar\epsilon{}^i\theta_i\right) \left(
\Psi^a\bPsi_a\right). \ee Therefore the full action \p{la8} will
be invariant only for $\alpha=-1$ which corresponds just to the
$SU(1,1|2)$ group! Clearly, with this value of $\alpha$ the first
term is also fixed to be ${\cal F}=X\log X$. As we already proved
this is not correct. The subtle point is the {\it locality} of the
transformation properties of $\Psi^a$. Indeed, from the explicit
form of $\delta \xi^a$ \p{ad7a}, it follows that they can be
explicitly integrated only for $\alpha=-1$. For any other value of
parameter $\alpha$ the integrated $\delta \xi^a$ which is just
$\delta \psi^a$ will contain non-local term. Thus, we see that
similarly to the preceding Sections the action does not seem to be
conformally invariant, but it possesses this invariance by
modification of the transformation properties of the involved
fields.

To conclude, let us make several comments.

Funny enough, but in contrast with the standard action \p{la5}, by
fixing the bosonic potential in the action \p{la8} to be $1/y^2$
in flat coordinates, one does not completely fix the prepotential
$F$. Indeed, rewriting \p{la8} in flat coordinates $y(x)$ with
$F''=(y')^2$ we get the condition
\be x\frac{d y}{dx} = a y
\quad \Rightarrow \quad y(x) = x^a.
\ee Thus, any polynomial superpotential ${\cal F}\sim X^a$, will give rise to a $N=4$ supersymmetric mechanics
with inverse square potential term in the bosonic sector.

One of the most interesting examples of such superpotentials are
the superconformally invariant ones \p{ad5} and \p{ad6}. Thus, the
actions of $D(2,1;\alpha)$ superconformal invariant mechanics read
\bea\label{AC1} S_{\alpha}&=& \int dt \left[ (1+\alpha)
\frac{{\dot y}{}^2}{2}+(1+\alpha)\frac{i}{8}\left(
\dot{\tilde\eta}{}^i\tilde{\bar\eta}{}_i-
\tilde\eta{}^i\dot{\bar{\tilde\eta}}_i\right)+\frac{i}{4}\left( {\dot w}_i {\bar w}{}^i-w_i\dot{\bar w}{}^i\right)-\frac{\alpha^2}{8(1+\alpha)y^2}\left(w^i{\bar w}_i\right)^2-\right.\nn\\
&& \left.\frac{\alpha}{8y^2}\tilde\eta_i\bar{\tilde\eta}_j\left( w^i {\bar w}{}^j+ w^j{\bar w}{}^i\right)+
\frac{(1+\alpha)(1+2\alpha)}{64 y^2} \tilde\eta{}^2 \bar{\tilde\eta}{}^2 \right], \qquad \alpha\neq -1,0
\eea
where
\be
y=x^{-\frac{1}{2\alpha}},\qquad \tilde\eta{}^i=x^{-\frac{1}{2\alpha}-1}\frac{\eta^i}{\alpha},
\ee
and
\bea\label{AC2}
S_{-1}&=& \int dt \left[  \frac{{\dot y}{}^2}{2}+\frac{i}{8}\left( \dot{\tilde\eta}{}^i\tilde{\bar\eta}{}_i-
\tilde\eta{}^i\dot{\bar{\tilde\eta}}_i\right)+\frac{i}{4}\left( {\dot w}_i {\bar w}{}^i-w_i\dot{\bar w}{}^i\right)-\frac{1}{8y^2}\left(w^i{\bar w}_i\right)^2-\right.\nn\\
&& \left.\frac{1}{8y^2}\tilde\eta_i\bar{\tilde\eta}_j\left( w^i
{\bar w}{}^j+ w^j{\bar w}{}^i\right)- \frac{1}{64 y^2}
\tilde\eta{}^2 \bar{\tilde\eta}{}^2 \right], \qquad \alpha=
-1,\quad y=\sqrt{x}, \; \tilde\eta{}^i=\frac{\eta^i}{\sqrt{x}}.
\eea Now it is clear that the simplest case of $N=4$
superconformal invariant mechanics corresponds just to the
$\alpha=-1/2$ case, i.e. the $OSp(4|2)$ superconformal group.
Indeed, it follows from \p{AC1}, that with $\alpha=-1/2$ the
four-fermionic interaction disappears from the Lagrangian. This
means that the corresponding supercharges contain the fermions
only linearly, similarly to the $N=2$ supersymmetric case.

Finally, one should note that our consideration in this Section is
very close to the one presented in the recent paper \cite{IFL}. Of
course, here we considered only the one-particle case and used the
standard $N=4, d=1$ superspace, in contrast with the harmonic
superspace approach \cite{HSS,DI,HSS1} advocated in \cite{IFL}. It
seems that our action \p{la8} is a more economical one. In any
case, the final component action \p{la10} has to coincide with the
one which could be obtained from the harmonic superspace action
presented in \cite{IFL} upon gauge fixing, integration over
theta's and harmonics, and elimination of the auxiliary
components.

\section{Conclusion}
In this paper we demonstrated that (super)conformal mechanics with
an additional harmonic oscillator term in the bosonic sector
possesses the same superconformal symmetry as the standard one.
The main difference between systems with and without oscillator
term is the modification of the transformation laws in such a way
that the kinetic term is invariant under the (super)conformal
group only together with the oscillator potential. The treatment
of the bosonic case has natural and straightforward extensions to
$N=2$ and $N=4$ superconformal symmetry. Probably, our approach
could be also extended to the case of $N$-extended supersymmetric
mechanics with $Osp(1,1|N/2)$ superconformal group \cite{IKL1}.
Another interesting question for further investigation is the
generalization of this approach to the $n$-particles
superconformal mechanics \cite{Cal}.

We also analyzed in this paper the generation of the bosonic
potentials in $N=4$ supersymmetric mechanics through coupling it
with auxiliary fermionic supermultiplet.  In contrast with
\cite{DI} the coupling we introduced does not advocate new
fermionic degrees of freedom - all our additional fermions are
purely auxiliary ones. The additional bosonic components have a
first order kinetic term and therefore they serve as spin degrees
of freedom. The new coupling we introduced in this paper is
invariant under the full $D(2,1;\alpha)$ superconformal group.
This invariance is not evident, because the starting action
possesses only $SU(1,1|2)$ invariance, provided we limit ourselves
by considering {\it local} transformation properties of involved
superfields. The invariance under the $D(2,1;\alpha)$
superconformal group is achieved within {\it non-local}
transformations, which become local in terms of new variables.
Thus, similarly to the situation with oscillator type potentials,
the key ingredient for the construction of the most general $N=4$
superconformal mechanics is the modification of (super)conformal
transformations.

Our approach in this paper is very close to the one recently
proposed in \cite{IFL}. It would be interesting to compare our
action with the one-particle action (with all fermionic terms
included) presented in \cite{IFL}. They have to coincide, because
the main ingredient - the action describing the coupling of the
basic $(1,4,3)$ supermultiplet $X$ with the auxiliary bosonic
$(4,4,0)$ one \p{ad3}- is unique and is completely fixed by $N=4$
Poincar\'{e} supersymmetry (up to an overall constant). The full
component action we constructed \p{AC1} revealed the main
peculiarity of the special case of $OSp(4|2)$ invariant action
which is the main subject considered in \cite{IFL}. In this
particular case the Lagrangian does not contain four-fermionic
interactions and, therefore, the corresponding supercharges are
linear over fermionic components, similarly to the case of $N=2$
supersymmetry.

Finally, the way to deal with spin degrees of freedom proposed in
the present work could be relevant for a proper supersymmetric
generalization of the system with Yang monopole recently analyzed
in \cite{toppp}.

\section*{Acknowledgments}
We acknowledge discussions with A.~Shcherbakov.

S.K. thanks the Laboratori Nazionali di Frascati
for the warm hospitality extended to
them during the course of this work.
This work was partially supported by INTAS under contract 05-7928,
RFBR grants 08-02-90490-Ukr, 06-02-16684 and  DFG grant 436 Rus~113/669/03.

\setcounter{equation}0
\def\theequation{A.\arabic{equation}}
\section*{Appendix. N=4, d=1 Superconformal algebra}
The most general $N{=}4, d{=}1$ superconformal algebra is the
superalgebra $D(2,1;\alpha)$.  We use the standard definition of
this superalgebra  \cite{FRS} with the notations of refs.
\cite{{IKLe},{IKLe1}}. It contains nine bosonic generators which
form a direct sum of $sl(2)$ with generators $P,D,K$ and two
$su(2)$ subalgebras with generators $V, \bV, V_3\, \; \mbox{ and }
\; T, \bT, T_3$, respectively: \bea\label{alg1} && i\left[
D,P\right] =P,\;  i\left[ D,K\right]=-K ,\; i\left[ P,K\right]=-2D
, \quad
i\left[ V_3,V\right]=-V,\;  i\left[ V_3,\bV \right]=\bV, \nn\\
&& i\left[ V,\bV\right]=2V_3,\quad
i\left[ T_3,T\right]=-T,\;  i\left[ T_3,\bT \right]=\bT,\;
i\left[ T,\bT\right]=2T_3.
\eea
The eight fermionic generators $Q^i,\bQ_i,S^i,\bS_i$
are in the fundamental representations of all bosonic subalgebras
(in our notation only
one $su(2)$ is manifest):
\bea\label{alg2}
&&i\left[D ,Q^i \right] = \frac{1}{2}Q^i,\;
i\left[D ,S^i \right] = -\frac{1}{2}S^i, \quad
i\left[P ,S^i \right] =-Q^i,\;
i\left[K ,Q^i \right] =S^i, \nonumber \\
&& i\left[V_3 ,Q^1 \right] =\frac{1}{2}Q^1,\; i\left[V_3 ,Q^2 \right]
=-\frac{1}{2}Q^2,\quad
i\left[V ,Q^1 \right] =Q^2, \; i\left[V ,\bQ_2 \right] =-\bQ_1, \nonumber \\
&&i\left[V_3 ,S^1 \right] =\frac{1}{2}S^1,\;    i\left[V_3 ,S^2 \right]
=-\frac{1}{2}S^2, \quad
i\left[V ,S^1 \right] =S^2, \;    i\left[V ,\bS_2 \right] =-\bS_1,  \nonumber\\
&& i\left[T_3 ,Q^i\right] =\frac{1}{2}Q^i, \; i\left[T_3
,S^i\right] =\frac{1}{2}S^i, \quad i\left[T ,Q^i\right] =\bQ^i, \;
i\left[T ,S^i\right] =\bS^i. \eea
The  fermionic
generators $Q^i,\bQ_k$ together with $P$ form the $N=4, d=1$ super
Poincar\'e subalgebra, while $S^i,\bS_k $ generate superconformal
translations: \be\label{allg3} \left\{Q^i ,\bQ_j \right\} =
-2\delta^i_j P , \quad \left\{S^i ,\bS_j \right\} =-2\delta^i_j K
. \ee The non-trivial dependence of the superalgebra
$D(2,1;\alpha)$ on the parameter $\alpha$ manifests itself only in
the cross-anticommutators of the Poincar\'e and conformal
supercharges \bea\label{alg4} && \left\{ Q^i,S^j \right\}
=-2(1+\alpha )\epsilon^{ij} \bT , \; \left\{Q^1 ,\bS_2 \right\}
=2\alpha \bV ,\;\left\{Q^1 ,\bS_1 \right\} =-2D-2\alpha
V_3+2(1+\alpha)T_3 ,
               \nonumber \\
&& \left\{Q^2 ,\bS_1 \right\} =-2\alpha V, \;\left\{Q^2 ,\bS_2
\right\} =-2D +2\alpha V_3+2(1+\alpha)T_3. \eea  The
generators $P,D,K$ are chosen to be hermitian, and the remaining
ones obey the following conjugation rules: \be\label{conjug}
\left( T \right)^\dagger = \bT, \; \left( T_3\right)^\dagger =-T_3
, \; \left( V \right)^\dagger = \bV, \; \left( V_3\right)^\dagger
=-V_3 , \; \overline{\left( Q^i \right)}=\bQ_i,\; \overline{\left(
S^i \right)}=\bS_i. \ee

The parameter $\alpha $ is an arbitrary real number. At $\alpha = 0$ and
$\alpha = -1$ one of the
$su(2)$ algebras decouples and the superalgebra $su(1,1\vert 2)\oplus su(2)$
is recovered.


\begin{thebibliography}{99}
\bibitem{AFF} V.~De~Alfaro, S.~Fubini, G.~Furlan, Nuovo Cimento, A34 (1974) 569.
\bibitem{n2} V.~Akulov, A.~Pashnev, Teor. Mat. Fiz. 56 (1983) 344; \\
S.~Fubini, E.~Rabinovici, Nucl. Phys. B245 (1984) 17.
\bibitem{IKL1} E.~Ivanov, S.~Krivonos, V.~Leviant, J. Phys. A: Math. Gen. 22 (1989) 4201.
\bibitem{AP1} V.P. Akulov, Sultan Catto, Oktay Cebecioglu, A. Pashnev, Phys.Lett. B575 (2003) 137,
{\tt  arXiv:hep-th/0303134 };\\
V.P. Akulov, Sultan Catto, A.I. Pashnev, ``N=2 SuperTime Dependent
Oscillator and spontaneous Breaking of Supersymmetry'',{\tt
arXiv:hep-th/0409316}.
\bibitem{IKL0} E.~Ivanov, S.~Krivonos,V.~Leviant, J. Phys. A: Math. Gen. 22 (1989) 345.
\bibitem{DI} F. Delduc, E. Ivanov, Nucl.Phys. B770 (2007) 179, {\tt arXiv:hep-th/0611247}.
\bibitem{IFL} S.~Fedoruk, E.~Ivanov, O.~Lechtenfeld, ``Supersymmetric Calogero models by gauging'', {\tt  arXiv:0812.4276}.
\bibitem{IO} E.~Ivanov, V. Ogievetsky, Teor. Mat. Fiz. 25 (1975) 164.
\bibitem{GR} S.J.~Gates Jr., L.~Rana, Phys. Lett. B342 (1995) 132, {\tt arXive:hep-th/9410150}.
\bibitem{root}S.~Bellucci, S.~Krivonos, A.~Marrani, E.~Orazi, Phys. Rev. D73 (2006) 025011,{\tt arXiv:hep-th/0511249}.
\bibitem{SM} S.~Bellucci, S.~Krivonos,  Lect. Notes Phys. 698 (2006) 49, {\tt arXiv:hep-th/0602199}.
\bibitem{IKLe1} E. Ivanov, S. Krivonos, O. Lechtenfeld, Class. Quant. Grav. 21 (2004) 1031,
{\tt arXiv:hep-th/0310299}.
\bibitem{HSS} A.S.~Galperin, E.A.~Ivanov, V.I.~Ogievetsky, E.S.~Sokatchev, {\it Harmonic superspace}, Cambridge University Press, 2001, 306p.
\bibitem{HSS1} E.~Ivanov, O.~Lechtenfeld, JHEP 0309 (2003) 073, {\tt arXiv:hep-th/0307111}.
\bibitem{Cal} S.~Bellucci, S.~Krivonos, A.~Sutulin, Nucl. Phys. B805 (2008) 24, {\tt arXiv:0805.3480}.
\bibitem{toppp}M.~Gonzales, Z.~Kuznetsova, A.~Nersessian, F.~Toppan, V.~Yeghikyan,
``Second Hopf map and supersymmetric mechanics with Yang
monopole'', {\tt  arXiv:0902.2682};\\
S. Bellucci, F. Toppan, V. Yeghikyan, ''Second Hopf map and
Yang-Coulomb system on 5d (pseudo)sphere'',  {\tt
arXiv:0905.3461}.
\bibitem{FRS} L. Frappat, A. Sciarrino, P. Sorba, ``Dictionary on Lie Superalgebras'', {\tt  arXiv:hep-th/9607161}.
\bibitem{IKLe} E. Ivanov, S. Krivonos, O. Lechtenfeld, JHEP 0303 (2003) 014,
{\tt arXiv:hep-th/0212303}.
\end{thebibliography}
\end{document}